\documentclass[a4paper,12pt]{article}
\usepackage[latin1]{inputenc}
\usepackage{amsfonts}
\usepackage{amsmath}
\usepackage{amssymb}

\input{tcilatex}

\begin{document}

\title{\flushright{\small
                   Brown preprint: BROWN-HET-1433} \bigskip \bigskip \\
\center{{\bf{Maps between Deformed and Ordinary Gauge Fields}}}}
\author{L. Mesref\thanks{%
On leave of absence from Département d'électrotechnique, Faculté de Génie
Electrique, U.S.T.O.M.B., Oran, Algeria.}}
\date{Department of Physics, Brown University, \\
Providence, Rhode Island 02912, USA.}
\maketitle

\begin{abstract}
In this paper, we introduce a map between the $q$-deformed gauge fields
defined on the GL$_{q}\left( N\right) $-covariant quantum hyperplane and the
ordinary gauge fields. Perturbative analysis of the $q$-deformed QED at the
classical level is presented and gauge fixing $\grave{a} $ la BRST is
discussed. An other star product defined on the hybrid $\left( q,h\right) $%
-plane is explicitly constructed .
\end{abstract}

\thispagestyle{empty}

\bigskip

\bigskip


Gerstenhaber star product, Jordanian deformation. \newline


\newpage

\section{Introduction}

Motivated by the need to control the divergences which occur in quantum
electrodynamics, Snyder \cite{snyder} proposed that one may use a
noncommutative structure of spacetime coordinates. Although its great
success, this suggestion has been swiftly forsaken. This is partly due to a
growing development in the renormalization program which captivated all the
attention of the leading physicists. The renormalization prescription solved
the quantum inconsistencies without making any ad hoc assumption on the
spacetime structure. Thanks to the seminal paper of Connes \cite{connes1}
the interest in noncommutativity \cite{connes2} has been revived. Natural
candidates for noncommutativity are provided by quantum groups \cite%
{drinfeld} which play the role of symmetry groups in quantum gauge theories 
\cite{gauge} (for a recent review see Ref. \cite{mesref0}). \newline
In Ref. \cite{mesref1} we have constructed a new map which relates a $q$%
-deformed gauge field defined on the Manin plane $\hat{x}\hat{y}=q\hat{y}%
\hat{x}$ and the ordinary gauge field. This map is the $q$-deformed analogue
of the Seiberg-Witten map \cite{seiberg}. We have found this map using the
Gerstenhaber star product \cite{gerstenhaber} instead of the
Groenewold-Moyal star product \cite{groenowold}. In the present paper, we
extend our analysis to the general GL$_{q}\left( N\right) $-covariant
quantum hyperplane defined by $\hat{x}^{i}\hat{x}^{j}=q\hat{x}^{j}\hat{x}%
^{i}\quad {\small {i<j}}$ and to the hybrid plane defined by $\hat{x}\hat{y}%
-q\hat{y}\hat{x}=h\hat{y}^{2}$.

\section{$q$-deformed gauge symmetry versus ordinary gauge symmetry}

To begin we consider the undeformed action 
\begin{equation}
S=\int d^{4}x\,\,\left[ \bar{\psi }\left( i\gamma ^{\mu }D_ {\mu }-m\right)
\psi -\frac{1}{4} F_{\mu \nu }F^{\mu \nu }\right],
\end{equation}
where

\begin{eqnarray}
D_{\mu }\psi &=& \left( \partial _{\mu } - i A_{\mu }\right) \psi  \notag \\
F_{\mu \nu } &=& \partial _{\mu }A_{\nu }-\partial _{\nu }A_{\mu }.
\end{eqnarray}

$S$ is invariant with respect to infinitesimal gauge transformations:

\begin{eqnarray}
\delta _{\lambda }A_{\mu }&=&\partial _{\mu }\lambda  \notag \\
\delta _{\lambda }\psi &=&i \lambda \psi  \notag \\
\delta _{\lambda }\bar{\psi }&=&-i \bar{\psi }\lambda .
\end{eqnarray}

Now let us study the quantum gauge theory on the quantum hyperplane $\hat{x}
^{i}\hat{x} ^{j}=q\hat{x} ^{j}\hat{x} ^{i} \quad i<j \quad q\in \mathbb{C}$. 
\newline
In general, the product of functions on a deformed space is defined via the
Gerstenhaber star product \cite{gerstenhaber}: Let $\mathcal{A}$ be an
associative algebra and let $D_{i},E^{i}:\mathcal{A}\rightarrow \mathcal{A}$
be a pairwise commuting derivations. \newline
Then the star product of $a$ and $b$ is given by

\begin{equation}
a\star b=\mu \circ e^{\zeta \sum_{i}D_{i}\otimes E^{i}} \left( a\otimes
b\right) ,
\end{equation}

where $\zeta $ is a parameter and $\mu $ the undeformed product given by

\begin{equation}
\mu \left( f\otimes g\right) =fg.
\end{equation}

On the quantum hyperplane $\hat{x} ^{i}\hat{x} ^{j}=q\hat{x} ^{j}\hat{x}
^{i} \quad i<j$, we can write this star product as:

\begin{equation}
f\star g=\mu \circ e^{\frac{i\eta }{2}\left( x^{i}\frac{\partial }{\partial
x^{i}}\otimes x^{j}\frac{\partial }{\partial x^{j}}-x^{j}\frac{\partial }{%
\partial x^{j}}\otimes x^{i}\frac{\partial }{\partial x^{i}}\right) } \left(
f\otimes g\right) .
\end{equation}

Let us note that the coordinates $x^{i}$ are commuting variables, while the
quantum coordinates $\widehat{x}^{i}$ are noncommuting variables. The
noncommutative algebra on the quantum hyperplane can be realized on the
algebra of the ordinary plane by using the Gerstenhaber star product.

A straightforward computation gives then the following commutation relations

\begin{equation}
x^{i}\star x^{j}=e^{\frac{i\eta }{2}}x^{i}x^{j},\qquad \qquad x^{j}\star
x^{i}=e^{\frac{-i\eta }{2} }x^{j}x^{i}.
\end{equation}

Whence

\begin{equation}
x^{i}\star x^{j}=e^{i\eta }x^{j}\star x^{i},\qquad q=e^{i\eta }.
\end{equation}

Thus we recover the commutation relations for the quantum hyperplane: 
\newline
$\hat{x}^{i}\hat{x}^{j}=q\hat{x}^{j}\hat{x}^{i}$. \newline
We can also write the product of functions as

\begin{equation}
f\star g=fe^{\frac{i}{2}\overleftarrow{\partial }_{k}\theta ^{kl}\left(
x\right) \overrightarrow{\partial }_{l}}g
\end{equation}

where $\theta ^{kl}\left( x\right) $ is an antisymmetric matrix depending on
the coordinates.

Expanding to first nontrivial order in $\eta $, we find

\begin{eqnarray}
f\star g &=& fg+\frac{i}{2}\eta x^{i} x^{j}\left( \frac{\partial f}{\partial
x^{i}}\frac{\partial g}{\partial x^{j}}-\frac{\partial f}{\partial x^{j}}%
\frac{\partial g}{\partial x^{i}}\right) +\circ \left( \eta ^{2}\right)
\qquad i<j  \notag \\
&=& fg+ \frac{i}{2}\theta ^{ij} \left( x\right) \partial _{i}f\,\,\partial
_{j}g +\circ \left( \eta ^{2}\right) .
\end{eqnarray}

The $q$-deformed infinitesimal gauge transformations are defined by

\begin{eqnarray}
\widehat{\delta }_{\widehat{\lambda }}\widehat{A}_{\mu } &=&\partial _{\mu }%
\widehat{\lambda }+i\left[ \widehat{\lambda },\widehat{A}_{\mu }\right]
_{\star }=\partial _{\mu }\widehat{\lambda }+i\widehat{\lambda }\star 
\widehat{A}_{\mu }-i\widehat{A}_{\mu }\star \widehat{\lambda },  \notag \\
\widehat{\delta }_{\widehat{\lambda }}\widehat{\psi }&=&i\widehat{\lambda }%
\star \widehat{\psi },  \notag \\
\widehat{\delta }_{\widehat{\lambda }}\widehat{\bar{\psi }}&=&-i \widehat{%
\bar{\psi }}\star \widehat{\lambda },  \notag \\
\widehat{\delta }_{\widehat{\lambda }}\widehat{F}_{\mu \nu } &=&i\widehat{%
\lambda }\star \widehat{F}_{\mu \nu }-i\widehat{F}_{\mu \nu }\star \widehat{%
\lambda }.
\end{eqnarray}

To first order in $\eta $, the above formulas for the gauge transformations
read

\begin{eqnarray}
\widehat{\delta }_{\widehat{\lambda }}\widehat{A}_{\mu } &=&\partial _{\mu }%
\widehat{\lambda }-\theta ^{\rho \sigma }\left( x\right) \partial _{\rho }%
\widehat{\lambda }\,\partial _{\sigma }\widehat{A}_{\mu } + \circ \left(
\eta ^{2}\right) ,  \notag \\
\widehat{\delta }_{\widehat{\lambda }}\widehat{\psi } &=& i\widehat{\lambda }
\widehat{\psi }-\theta ^{\rho \sigma }\left( x\right) \partial _{\rho }%
\widehat{\lambda }\,\partial _{\sigma } \widehat{\psi }+ \circ \left( \eta
^{2}\right) ,  \notag \\
\widehat{\delta }_{\widehat{\lambda }}\widehat{\bar{\psi }} &=& -i\widehat{%
\lambda } \widehat{\bar{\psi }}+\theta ^{\rho \sigma }\left( x\right)
\partial _{\rho }\widehat{\bar{\psi }}\,\partial _{\sigma } \widehat{\lambda 
}+ \circ \left( \eta ^{2}\right) ,  \notag \\
\widehat{\delta }_{\widehat{\lambda }}\widehat{F}_{\mu \nu } &=&-\theta
^{\rho \sigma }\left( x\right) \partial _{\rho }\widehat{\lambda } \partial
_{\sigma }\,\widehat{F}_{\mu \nu } + \circ \left( \eta ^{2}\right) .
\end{eqnarray}

To ensure that an ordinary gauge transformation of $A$ by $\lambda $ is
equivalent to $q$-deformed gauge transformation of $\widehat{A}$ by $%
\widehat{\lambda }$ we consider the following relation \cite{seiberg}

\begin{equation}
\widehat{A}\left( A\right) +\widehat{\delta }_{\widehat{\lambda }}\widehat{A}%
\left( A\right) =\widehat{A}\left( A+\delta _{\lambda }A\right) .
\end{equation}

We first work the first order in $\theta $

\begin{eqnarray}
\widehat{A} &=&A+A^{\prime }\left( A\right)  \notag \\
\widehat{\lambda }\left( \lambda ,A\right) &=&\lambda +\lambda ^{\prime
}\left( \lambda ,A\right) .
\end{eqnarray}

Expanding in powers of $\theta $ we find

\begin{equation}
A_{\mu }^{\prime }\left( A+\delta _{\lambda }A\right) -A_{\mu }^{\prime
}\left( A\right) -\partial _{\mu }\lambda ^{\prime }=\theta ^{kl}\left(
x\right) \partial _{k}A_{\mu }\partial _{l}\lambda .
\end{equation}

The solutions are given by

\begin{eqnarray}
\widehat{A}_{\mu } &=&A_{\mu }-\frac{1}{2}\theta ^{\rho \sigma }\left(
x\right) \left( A_{\rho }F_{\sigma \mu }+A_{\rho }\partial _{\sigma }A_{\mu
}\right) , \\
\widehat{\lambda } &=&\lambda +\frac{1}{2}\theta ^{\rho \sigma }\left(
x\right) A_{\sigma }\partial _{\rho }\lambda  \notag \\
\widehat{\psi } &=&\psi +\frac{1}{2}\theta ^{\rho \sigma }\left( x\right)
A_{\sigma }\partial _{\rho }\psi .
\end{eqnarray}

The $q$-deformed curvature $\widehat{F}_{\mu \nu }$ is given by

\begin{eqnarray}
\widehat{F}_{\mu \nu } &=&\partial _{\mu }\widehat{A}_{\nu }-\partial _{\nu }%
\widehat{A}_{\mu }-i\left[ \widehat{A}_{\mu },\widehat{A}_{\nu }\right]
_{\star }  \notag \\
&=&\partial _{\mu }\widehat{A}_{\nu }-\partial _{\nu }\widehat{A}_{\mu }-i%
\widehat{A}_{\mu }\star \widehat{A}_{\nu }+i\widehat{A}_{\nu }\star \widehat{%
A}_{\mu }.
\end{eqnarray}

Finally, we find

\begin{eqnarray}
\widehat{F}_{\mu \nu } &=&F_{\mu \nu }+\theta ^{\rho \sigma }\left( x\right)
\left( F_{\mu \rho }F_{\nu \sigma }-A_{\rho }\partial _{\sigma }F_{\mu \nu
}\right)  \notag \\
&&-\frac{1}{2}\partial _{\mu }\theta ^{\rho \sigma }\left( x\right) \left(
A_{\rho }F_{\sigma \nu }+A_{\rho }\partial _{\sigma }A_{\nu }\right)  \notag
\\
&&+\frac{1}{2}\partial _{\nu }\theta ^{\rho \sigma }\left( x\right) \left(
A_{\rho }F_{\sigma \mu }+A_{\rho }\partial _{\sigma }A_{\mu }\right) .
\end{eqnarray}

From this equation we can see the appearance of terms proportional to $%
\partial _{\mu }\theta ^{\rho \sigma }\left( x\right) $. Equation (19) can
also be written as

\begin{equation}
\widehat{F}_{\mu \nu }=F_{\mu \nu }+f_{\mu \nu }+o\left( \eta ^{2}\right) ,
\end{equation}

where $f_{\mu \nu }$ is the quantum correction linear in $\eta $. The
quantum analogue of Equ. (1) is given by

\begin{equation}
\hat{S}=\int d^{4}x\,\,\left[ \hat{\bar{\psi }}\star \left( i\gamma ^{\mu }%
\hat{D}_ {\mu }-m\right) \hat{\psi }-\frac{1}{4} \hat{F}_{\mu \nu }\star 
\hat{F}^{\mu \nu }\right] ,
\end{equation}

where $\hat{D}_{\mu }\hat{\psi }=\partial _{\mu }\hat{\psi }-i\hat{A}_{\mu
}\star \hat{\psi }$.

We can easily see from this equation that the $q$-deformed action contains
non-renormalizable vertices of dimension six. Other terms which are
proportional to $\partial _{\mu }\theta ^{\rho \sigma }\left( x\right) $
appear. \newline
A gauge fixing term is needed in order to quantize the system. This is done
in the BRST and anti-BRST formalism. As usual, the BRST transformations are
obtained by replacing $\hat{\lambda }$ by $\hat{c}$ and are given by:

\begin{eqnarray}
\widehat{s}\widehat{A}_{\mu } &=&\partial _{\mu }\widehat{c}-\theta ^{\rho
\sigma }\left( x\right) \partial _{\rho }\widehat{c}\,\partial _{\sigma }%
\widehat{A}_{\mu }+\circ \left( \eta ^{2}\right) ,  \notag \\
\widehat{s}\widehat{\psi } &=&i\widehat{c}\,\widehat{\psi }-\theta ^{\rho
\sigma }\left( x\right) \partial _{\rho }\widehat{c}\,\partial _{\sigma }%
\widehat{\psi }+\circ \left( \eta ^{2}\right) ,  \notag \\
\widehat{s}\widehat{\bar{\psi}} &=&-i\widehat{\bar{\psi}}\widehat{c}+\theta
^{\rho \sigma }\left( x\right) \partial _{\rho }\widehat{\bar{\psi}}%
\,\partial _{\sigma }\widehat{c}+\circ \left( \eta ^{2}\right) ,  \notag \\
\widehat{s}\widehat{F}_{\mu \nu } &=&-\theta ^{\rho \sigma }\left( x\right)
\partial _{\rho }\widehat{c}\,\partial _{\sigma }\widehat{F}_{\mu \nu
}+\circ \left( \eta ^{2}\right) ,  \notag \\
\hat{s}\hat{\bar{c}} &=&b,\quad \hat{s}\hat{c}=0,\quad \hat{s}\hat{b}=0,
\end{eqnarray}

where $\widehat{c}, \widehat{\bar{c}} $ are the quantum Faddeev-Popov ghost
and anti-ghost fields, $\widehat{b} $ a scalar field (sometimes called the
Nielson-Lautrup auxiliary field) and $\widehat{s} $ the quantum BRST
operator. The gauge-fixing term is introduced as

\begin{equation}
\widehat{S}_{\mathrm{gf}}+\int d^{4}x \quad \widehat{s} \,\, \left( \widehat{%
\bar{c}} \star \left( \frac{\alpha }{2} \widehat{b} - \partial _{\mu } 
\widehat{A}^{\mu } \right) \right) .
\end{equation}

An expansion in $\eta $ leads to an action corresponding to a highly
nonlinear gauge.

The external field contribution is given by

\begin{equation}
\widehat{S} _{\mathrm{ext}}= \int d^{4}x \quad \left( \widehat{A}^{\ast \mu
} \star \widehat{s} \widehat{A} _{\mu } + \widehat{c^{\ast }} \star \widehat{%
s} \widehat{c} \right) ,
\end{equation}

where $\widehat{A^{\star }},\,\widehat{c^{\star }}$ are external fields
(called antifields in the Batalin-Vilkovisky formalism) and play the role of
sources for the BRST- variations of the fields $\widehat{A},\,\widehat{c}$. 
\newline
The $\widehat{c}$ and $\widehat{\bar{c}}$ play quite asymmetric roles, they
cannot be related by Hermitian conjugation. The anti- BRST transformations
are given by

\begin{eqnarray}
\widehat{\bar{s}}\widehat{A}_{\mu } &=&\partial _{\mu }\widehat{\bar{c}}%
-\theta ^{\rho \sigma }\left( x\right) \partial _{\rho }\widehat{\bar{c}}%
\,\partial _{\sigma }\widehat{A}_{\mu },  \notag \\
\widehat{\bar{s}}\widehat{\psi } &=&i\widehat{\bar{c}}\widehat{\psi }-\theta
^{\rho \sigma }\left( x\right) \partial _{\rho }\widehat{\bar{c}}\,\partial
_{\sigma }\widehat{\psi }+\circ \left( \eta ^{2}\right) ,  \notag \\
\widehat{\bar{s}}\widehat{\bar{\psi}} &=&-i\widehat{\bar{\psi}}\,\widehat{%
\bar{c}}+\theta ^{\rho \sigma }\left( x\right) \partial _{\rho }\widehat{%
\bar{\psi}}\,\partial _{\sigma }\widehat{\bar{c}}+\circ \left( \eta
^{2}\right) ,  \notag \\
\widehat{\bar{s}}\widehat{F}_{\mu \nu } &=&-\theta ^{\rho \sigma }\left(
x\right) \partial _{\rho }\widehat{\bar{c}}\,\,\partial _{\sigma }\widehat{F}%
_{\mu \nu }+\circ \left( \eta ^{2}\right) ,  \notag \\
\widehat{\bar{s}}\widehat{\bar{c}} &=&0,\quad \widehat{\bar{s}}\widehat{c}%
=-b,\quad \widehat{\bar{s}}\widehat{b}=0.
\end{eqnarray}

Here $\widehat{\bar{s}}$ is the quantum anti- BRST operator. The complete
tree-level action is given by:

\begin{equation}
\Sigma \left( \widehat{A} _{\mu }, \widehat{c}, \widehat{\bar{c}}, \widehat{b%
}, \widehat{A^{\ast }}_{\mu }, \widehat{c^{\ast }} \right) = \widehat{S} + 
\widehat{S}_{\mathrm{gf}} + \widehat{S}_{\mathrm{ext}} .
\end{equation}

If we replace the ordinary fields and the ordinary action by their $q$%
-deformed analogues we can construct a $q$-deformed partition function. This
enables us to study the $q$-perturbative theory, find the $q$-deformed $n$%
-point correlation functions and defined the $q$-deformed analogue of the
Slavnov-Taylor identity.

\section{$\left( q,h\right) $-deformed gauge symmetry versus ordinary gauge
symmetry}

It is well known \cite{demidov, aghamohammadi} that the only quantum groups
which preserve nondegenerate bilinear forms are GL$_{qp}\left( 2\right) $
and GL$_{hh^{\prime }}\left( 2\right) $. They act on the $q$-plane (Manin
plane) defined by $\widehat{X}\widehat{Y}=q\widehat{Y}\widehat{X}$ and on
the $h$-plane (Jordanian plane) defined by $\widehat{x}\widehat{y}-\widehat{y%
} \widehat{x}=h\widehat{y^{2}}$, respectively. \newline
In this section we give the form of the star product of functions defined on
the hybrid plane $\left( q,h\right) $-plane defined by $\widehat{x}\widehat{y%
}-q\widehat{y} \widehat{x}=h\widehat{y^{2}}$. This map can be used to relate
the $\left( q,h\right) $-gauge fields to the ordinary ones. \newline

Let us recall that the Manin plane and the Jordanian plane are related by a
transformation \cite{aghamohammadi}

\begin{eqnarray}
\left( 
\begin{array}{c}
\widehat{X} \\ 
\widehat{Y}%
\end{array}
\right) &=& \left( 
\begin{array}{cc}
1 & \alpha \\ 
0 & 1%
\end{array}
\right) \left( 
\begin{array}{c}
\widehat{x} \\ 
\widehat{y}%
\end{array}
\right) ,  \notag \\
\left( 
\begin{array}{c}
\partial _{\widehat{X}} \\ 
\partial _{\widehat{Y}}%
\end{array}
\right) &=& \left( 
\begin{array}{cc}
1 & 0 \\ 
-\alpha & 1%
\end{array}
\right) \left( 
\begin{array}{c}
\partial _{\widehat{x}} \\ 
\partial _{\widehat{y}}%
\end{array}
\right) ,
\end{eqnarray}

where $\alpha =\frac{h}{q-1}$.

The star product of functions on the Manin plane is defined by choosing the
pairwise commuting derivations: $X \frac{\partial }{\partial X}$ and $Y 
\frac{\partial } {\partial Y}$ .

\begin{equation}
f \star g = \mu \circ e^{\frac{i \eta }{2}\left( X \frac{\partial }{\partial
X} \otimes Y \frac{\partial }{\partial Y}-Y \frac{\partial }{\partial Y}
\otimes X \frac{\partial }{\partial X} \right) } \left( f \otimes g \right) .
\end{equation}

A straightforward computation gives the following commutation relations

\begin{equation}
X \star Y = e^{\frac{i \eta }{2}}X Y, \qquad Y \star X = e^{\frac{-i \eta }{2%
}} Y X
\end{equation}

Whence

\begin{equation}
X \star Y = e^{i \eta } Y \star X, \qquad q = e^{i \eta }.
\end{equation}

Thus we recover the commutation relations for the Manin plane.

On the hybrid space, we define the star product as

\begin{equation}
f \star g = \mu \circ e^{\frac{i \eta }{2} \left[ \left( x\frac{\partial } {%
\partial x}+ \alpha y \frac{\partial }{\partial x}\right) \otimes \left( y 
\frac{\partial }{\partial y}- \alpha y \frac{\partial }{\partial x} \right)
- \left( y \frac{\partial }{\partial y} - \alpha y \frac{\partial }{\partial
x}\right) \otimes \left( x \frac{\partial }{\partial x} + \alpha y \frac{%
\partial } {\partial x} \right) \right] }\left( f \otimes g \right) .
\end{equation}

A direct computation gives

\begin{equation}
x\star y=e^{\frac{i\eta }{2}}xy+\left( e^{\frac{i\eta }{2}}-1\right) \alpha
y^{2},\quad y\star x=e^{-\frac{i\eta }{2}}yx+\left( e^{-\frac{i\eta }{2}%
}-1\right) \alpha y^{2}.
\end{equation}

Whence

\begin{eqnarray}
x\star y &=&e^{i\eta }y\star x+\left( e^{i\eta }-1\right) \alpha y^{2} 
\notag \\
&=&q\,y\star x+h\,y^{2}.
\end{eqnarray}

Thus we recover the commutation relations for the hybrid plane.

Expanding to first nontrivial order in $\eta $ and $h$ we find

\begin{equation}
f\star g=f\,g+\frac{i}{2}\left( \eta xy-ihy^{2}\right) \left( \frac{\partial
f}{\partial x}\frac{\partial g}{\partial y}-\frac{\partial f}{\partial y}%
\frac{\partial g}{\partial x}\right) .
\end{equation}

\bigskip

If we take $\eta =h = 0$ we recover the ordinary product of commuting
functions defined on the ordinary two-dimensional plane.

We can also write the star product\ (defined by Equ. (31) ) as

\begin{equation}
f \star g = f e^{\frac{i}{2} \overleftarrow{\partial _{k}} \Theta ^{k l}
\left( x,y\right) \overrightarrow{\partial _{l} }} g .
\end{equation}

\bigskip

Here the antisymmetric matrix $\Theta ^{k l}\left( x,y\right) = \left( \eta
x y -i h y^{2}\right) \epsilon ^{k l} $ with \newline
$\epsilon ^{1 2}= \epsilon ^{2 1}=1$.

This star product can be used to relate the $\left( q,h\right) $-deformed
gauge fields to the ordinary ones.

\paragraph{Acknowledgements.}

I would like to extend very special thanks to the members of the High Energy
Theory Group at Brown university for warm hospitality. This work was partly
supported by Université des sciences et de la Technologie d'Oran,
U.S.T.O.M.B., Oran, Algeria.

\end{document}